\documentclass[11pt]{article}
\usepackage{graphicx}
\usepackage{mathrsfs}
\usepackage{amssymb}	
\usepackage{amsmath}
\usepackage{xcolor}
\usepackage{placeins}
\usepackage[normalem]{ulem}
\usepackage{stmaryrd}

\usepackage[colorlinks=true,citecolor=blue]{hyperref}
\definecolor{lcolor}{rgb}{0.5,0,0}
\definecolor{citcolor}{rgb}{0,0.3,0.0}
\definecolor{coloryksi}{rgb}{0.5,0.0,0.0}
\definecolor{colorkaksi}{rgb}{0.0,0.0,0.5}
\definecolor{colorkolme}{rgb}{0.0,0.3,0.1}

\usepackage{mathtools}
\usepackage{physics}
\usepackage{tensor}
\usepackage{cancel}
\usepackage{mathrsfs}

\usepackage{mciteplus}

\usepackage{wrapfig}
\usepackage{tikz}
\usetikzlibrary{snakes}

\newcommand{\qqbg}{{q\Bar{q}g}}

\newcommand{\xpom}{{x_\mathbb{P}}}
\newcommand{\xbj}{{x_\mathrm{Bj}}}

\newcommand{\xt}{{\mathbf{x}}}
\newcommand{\cxt}{{\conj{\xt}}}

\newcommand{\Yt}{{\mathbf{Y}}}

\newcommand{\pt}{{\mathbf{p}}}

\newcommand{\Deltat}{{\boldsymbol{\Delta}}}

\newcommand{\besj}{{\mathrm{J}}}
\newcommand{\besk}{{\mathrm{K}}}

\newcommand{\ud}{\, \mathrm{d}}

\newcommand{\nc}{{N_\mathrm{c}}}

\newcommand{\half}{\frac{1}{2}}

\newcommand{\cf}{C_\mathrm{F}}

\newcommand{\as}{\alpha_{\mathrm{s}}}

\newcommand{\conj}[1]{\mathop{\overline{#1}}\nolimits}

\newcommand{\contrip}{\conj{0} \conj{1} \conj{2}}

\newcommand{\mx}{M_X}

\newcounter{diag}
\newcounter{subdiag}[diag]
\newcommand{\namediag}[1]{\refstepcounter{diag} \thediag \label{#1}}

\renewcommand{\thediag}{(\alph{diag})}

\newcommand\pubdate{\today}

\def\Title#1{\begin{center} {\Large #1 } \end{center}}
\def\Author#1{\begin{center}{ \sc #1} \end{center}}
\def\Address#1{\begin{center}{ \it #1} \end{center}}

\newcommand\pubblock{\rightline{\begin{tabular}{l}  \\ 
         \pubdate  \end{tabular}}}
\newenvironment{Abstract}{\begin{quotation}  }{\end{quotation}}
\newenvironment{Presented}{\begin{quotation} \begin{center} 
             PRESENTED AT\end{center}\bigskip 
      \begin{center}\begin{large}}{\end{large}\end{center} \end{quotation}}

\begin{document}
\pagenumbering{gobble}
\begin{titlepage}
 \pubblock
\vfill
\Title{Diffractive Deep Inelastic Scattering in the Dipole Picture at Next-to-Leading Order}
\vfill
\Author{G. Beuf}
\Address{National Centre for Nuclear Research, 02-093 Warsaw, Poland}
\Author{H. Hänninen}
\Address{ Department of Mathematics and Statistics, University of Jyv\"askyl\"a, \\
P.O. Box 35, 40014 University of Jyv\"asky\"a, Finland}
\Author{T. Lappi, H. Mäntysaari}
\Address{ Department of Physics, University of Jyv\"askyl\"a, \\
P.O. Box 35, 40014 University of Jyv\"asky\"a, Finland}
\Address{ Helsinki Institute of Physics, P.O. Box 64, 00014 University of Helsinki, Finland}
\Author{Y. Mulian}
\Address{Instituto Galego de Fisica de Altas Enerxias IGFAE, Universidade de Santiago de Compostela, 15782 Santiago de Compostela, Galicia-Spain}
\vfill
\begin{Abstract}
We calculate the contribution from the $q \bar q g$ state production to the diffractive cross sections in deep inelastic scattering at high energy. The obtained cross section is finite by itself, and consists a part of the full next-to-leading order result for the diffractive structure functions.
Our calculation for the diffractive structure functions is performed using exact kinematics, under the shockwave approximation of the scattering process. Once the calculation is completed, we show that the previously known behaviour at the high-$Q^2$ and large-$\mx^2$ regime can be extracted from our results by taking the appropriate limits.
Furthermore, we discuss the steps required to obtain the complete next-to-leading order results for the structure functions in the color glass condensate (CGC) formalism, and the application of these results to phenomenology.
\end{Abstract}
\vfill
\newpage
\begin{Presented}
DIS2023: XXX International Workshop on Deep-Inelastic Scattering and
Related Subjects, \\
Michigan State University, USA, 27-31 March 2023 \\
     \includegraphics[width=9cm]{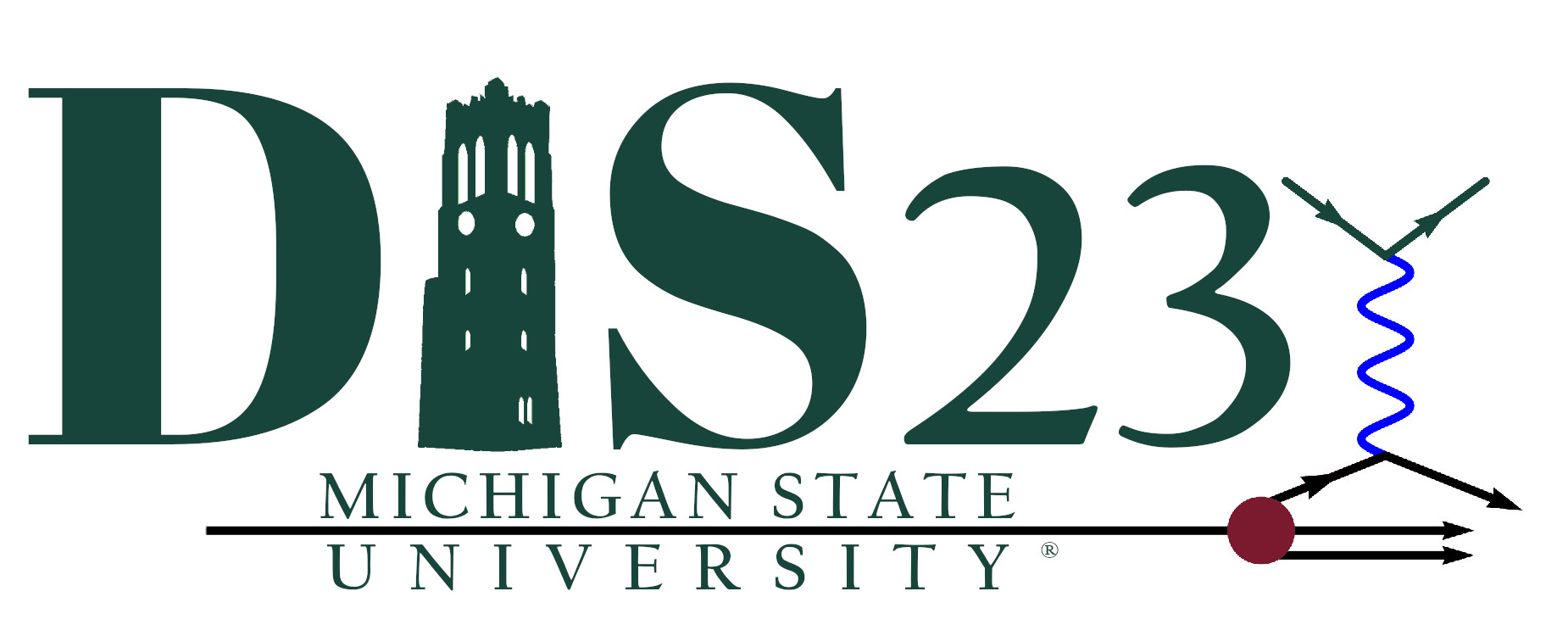}
\end{Presented}
\vfill
\end{titlepage}

\section{Introduction}
\pagenumbering{arabic}
    During the early 1990s, when the HERA particle collider at DESY started its operation, a remarkable observation was made: the target protons would survive intact in a large fraction of collisions with the probe lepton, and a jet would be seen in the detector. In these events a large separation was observed between the outgoing proton and the jet, making the scattering diffractive --- strongly interacting but without net color charge exchange.

    Diffractive deep inelastic scattering (DDIS) can be naturally described in the dipole picture and color glass condensate effective field theory. In this article we highlight our recent next-to-leading order accuracy calculation of the $\qqbg$ contribution to the diffractive structure functions \cite{Beuf:2022kyp}.

\section{Diffractive DIS in the Dipole Picture}

\begin{wrapfigure}{r}{7cm}
\includegraphics[width=6.00cm]{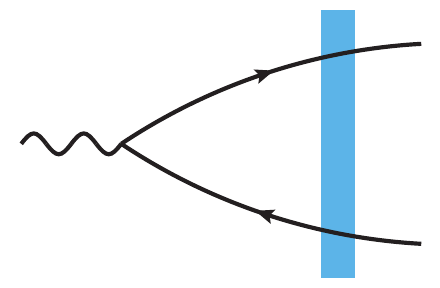}
\begin{tikzpicture}[overlay]
 \node[anchor=south east] at (-5cm,2.1cm) {$q,\lambda$};
 \node[anchor=south west] at (-1.2cm,2.6cm) {$z_0, \pt_0, \xt_0$};
  \node[anchor=south west] at (-1.2cm,0.8cm) {$z_1, \pt_1, \xt_1$};
\end{tikzpicture}
\caption{The leading order $\gamma^* p$ diffractive DIS amplitude. The blue band represents the interaction with the target color field.}
\label{fig:ddis_loampli}
\end{wrapfigure}
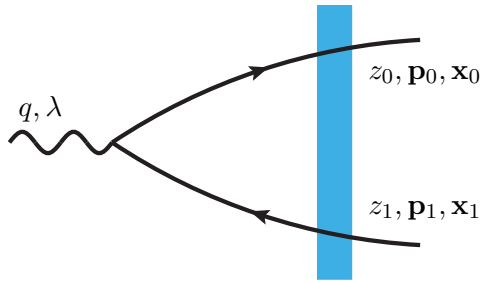

    Diffractive DIS in our approach looks very alike to inclusive DIS: first the virtual photon fluctuates into a quark-antiquark pair, which then scatters off the target color-field as shown in Fig. \ref{fig:ddis_loampli}. The diffractive process is then imposed by specifying that the scattered outgoing Fock-state is a color singlet, which ensures that there is no net color exchange in the scattering.
    By the time that HERA began its operation the result for the diffractive structure function at the leading order accuracy was already known~\cite{Nikolaev:1991et}.

    The necessity of including the transverse $\qqbg$ contribution in order to describe $F_2^{\mathrm{D}}$ at low-$\beta$ was observed in Ref.~\cite{Golec-Biernat:1999qd}, as the quark-antiquark contribution vanishes as $\beta \to 0$. This challenging next-to-leading order contribution was calculated under some simplifying assumptions, such as the large-$\mx$ limit \cite{Munier:2003zb}, and the large-$Q^2$ limit \cite{Golec-Biernat:1999qd,Wusthoff:1997fz}. The large-$Q^2$ limit result lead to the seminal work by Golec-Biernat and Wusthoff \cite{Golec-Biernat:1999qd}, where they were able to describe the diffractive measurements using their dipole scattering parametrization based on DIS data.

    In \cite{Beuf:2022kyp}, our first step was to re-derive the leading order $q \bar q$ results using the modern formalism of light-cone wavefunctions (LCWF), and keeping the momentum transfer $t$ dependence in anticipation of data from the planned EIC experiments. The $t$ and $\mx$ dependencies factorize for the structure functions even in the general case, leading to neatly contained factors related to the diffractive production of the final state with invariant mass $\mx$ at momentum transfer $t$:
    \begin{equation}
    \label{eq:I2_Delta_result}
        {\cal I}_\Deltat^{(2)} = \frac{1}{4\pi} \besj_0\left( \sqrt{\abs{t}}\: \norm{z_0 \xt_{\Bar 0 0} - z_1\xt_{\Bar 1 1} } \right),
    \end{equation}
    \begin{equation}
    \label{eq:IMX_2}
        {\cal I}_{\mx}^{(2)} = 
        \frac{1}{4\pi} 
        \besj_0\left( \sqrt{z_0 z_1}\mx \norm{\xt_{\Bar 0 \Bar 1} - \xt_{01} } \right),
    \end{equation}
    where the superscript $(2)$ refers to the number of partons in the final state.
    
    The factorization is analogous for the 3-parton Fock-state, however, the calculations are more involved:
    \begin{equation}
        {\cal I}_\Deltat^{(3)} = \frac{1}{4 \pi} \mathrm{J}_0 \left( \sqrt{\abs{t}} \norm{z_0 \xt_{\Bar 0 0} + z_1 \xt_{\Bar 1 1} + z_2 \xt_{\Bar 2 2}} \right),
    \end{equation}
    \begin{align}
    \label{eq:IMX3}
        {\cal I}_{\mx}^{(3)} =
            2 \frac{z_0 z_1 z_2}{(4 \pi)^2} \frac{\mx}{Y_{012}} \mathrm{J}_1 \left( \mx Y_{012} \right),
    \end{align}
    \begin{equation}
        \Yt_{012}^2 =
            z_0 z_1 \left( \xt_{\Bar 0 0} - \xt_{\Bar 1 1} \right)^2 +
            z_1 z_2 \left( \xt_{\Bar 2 2} - \xt_{\Bar 1 1} \right)^2 +
            z_0 z_2 \left( \xt_{\Bar 2 2} - \xt_{\Bar 0 0} \right)^2.
    \end{equation}

\section{Outline of the NLO calculation and results}

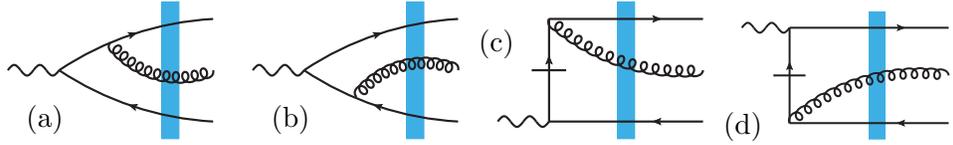
\begin{figure*}[tb!]
    \centerline{
    \includegraphics[width=3cm]{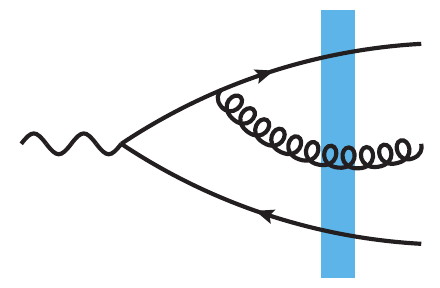}
    \begin{tikzpicture}[overlay]
    \node[anchor=south west] at (-3cm,0cm) {\namediag{diag:ddis_gwavef1}};
    \end{tikzpicture}
    \includegraphics[width=3cm]{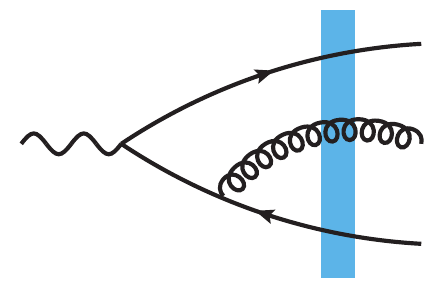}
    \begin{tikzpicture}[overlay]
    \node[anchor=south west] at (-3cm,0cm) {\namediag{diag:ddis_gwavef2}};
    \end{tikzpicture}
    \includegraphics[width=3cm]{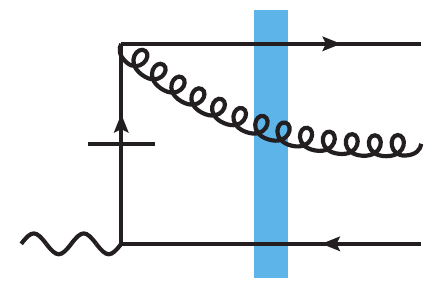}
    \begin{tikzpicture}[overlay]
    \node[anchor=south west] at (-3.5cm,1cm) {\namediag{diag:ddis_gwavefinst1}};
    \end{tikzpicture}
    \includegraphics[width=3cm]{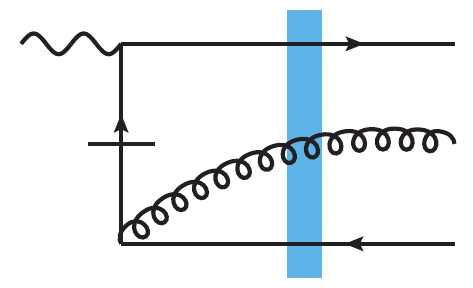}
    \begin{tikzpicture}[overlay]
    \node[anchor=south west] at (-3.5cm,-0.1cm) {\namediag{diag:ddis_gwavefinst2}};
    \end{tikzpicture}
    }
    \caption{
    A part of the contributions for the $\qqbg$ production: gluon emission diagrams where the gluon crosses the shockwave and is produced in the final state.
    }\label{fig:ddis_nlo}
\end{figure*}

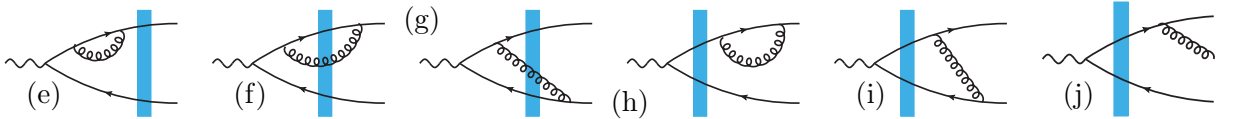
\begin{figure*}[tb!]
    \centerline{
    \includegraphics[width=2.5cm]{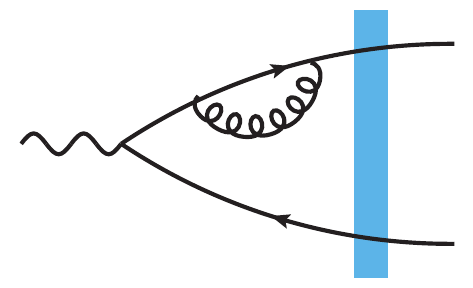}
    \begin{tikzpicture}[overlay]
    \node[anchor=south west] at (-2.5cm,0cm) {\namediag{diag:ddis_prop1ampli}};
    \end{tikzpicture}
    \includegraphics[width=2.5cm]{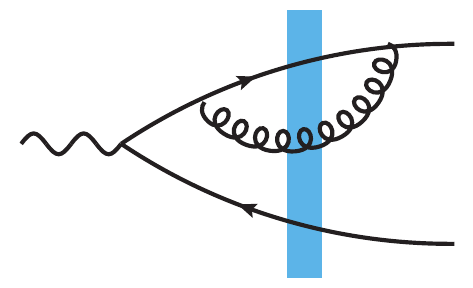}
    \begin{tikzpicture}[overlay]
    \node[anchor=south west] at (-2.5cm,0cm) {\namediag{diag:ddis_propshock1ampli}};
    \end{tikzpicture}
    \includegraphics[width=2.5cm]{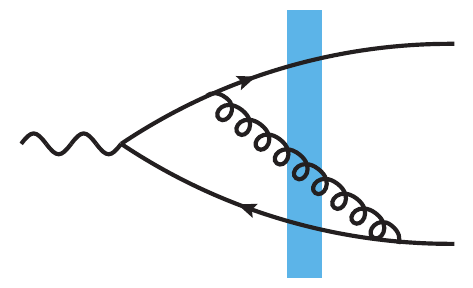}
    \begin{tikzpicture}[overlay]
    \node[anchor=south west] at (-3cm,1cm) {\namediag{diag:ddis_vertexshock1ampli}};
    \end{tikzpicture}
    \includegraphics[width=2.5cm]{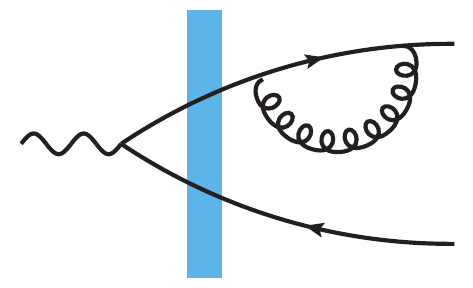}
    \begin{tikzpicture}[overlay]
    \node[anchor=south west] at (-3cm,-0.1cm) {\namediag{diag:ddis_propfs1ampli}};
    \end{tikzpicture}
    \includegraphics[width=2.5cm]{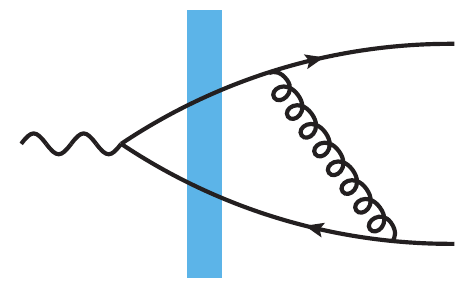}
    \begin{tikzpicture}[overlay]
    \node[anchor=south west] at (-2.5cm,0cm) {\namediag{diag:vertexfs1}};
    \end{tikzpicture}
    \includegraphics[width=2.5cm]{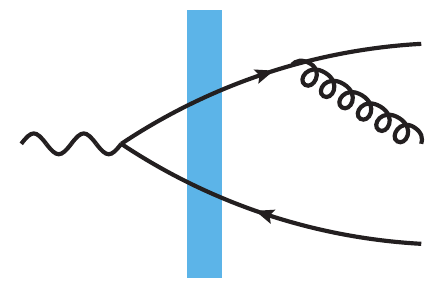}
    \begin{tikzpicture}[overlay]
    \node[anchor=south west] at (-2.5cm,0cm) {\namediag{diag:ddis_gwaveffs1}};
    \end{tikzpicture}
    }
    \caption{
    Gluon emission diagrams that are not included in the present results. These contributions include gluon loop corrections before the shock, through the shock and after the shock. The gluon emission can also take place in the final state and the emitted gluon can either be produced or re-absorbed in the final state.
    }\label{fig:ddis_loop}
\end{figure*}

Leveraging the theory work already laid out for DIS at NLO \cite{Beuf:2016wdz,Beuf:2017bpd}, the calculation of the $\qqbg$ contribution to the diffractive structure functions proceeds largely in the same vein as the leading order case discussed above. The diffractive final state is a Fock-state of 3 partons as shown in Fig. \ref{fig:ddis_nlo}, and its formation is given by the virtual photon LCWF $\gamma^* \to \qqbg$, which was calculated in Refs. \cite{Beuf:2016wdz,Beuf:2017bpd}. The calculation of the $\qqbg$ contribution then amounts to multiplying the $\gamma^* \to \qqbg$ amplitude by its complex conjugate, and combining the result with the calculation of the 3-parton Fock-state scattering from the target color field, which we performed in \cite{Beuf:2022kyp}.
This results in a finite and physical cross section for $\qqbg$ production. The remaining loop and final state radiation contributions (Fig. \ref{fig:ddis_loop}) are left for future work~\cite{Penttala:2023}.
More detailed, perhaps more pedagogical, steps are found in Ref.~\cite{Hanninen:2021byo}. The partial $\qqbg$ contributions to the diffractive structure functions are
\begin{align}
	\notag
	& \xpom F_{L, \, q \Bar q  g}^{\textrm{D}(4) \, \textrm{NLO}} (\xbj, Q^2, \beta, t)
	=
	4
	\frac{\as \nc \cf Q^4 }{\beta}
	\sum_f e_f^2
	\int_0^1 \frac{\ud z_0}{z_0}
	\int_0^1 \frac{\ud z_1}{z_1}
	\int_0^1 \frac{\ud z_2}{z_2}
	\delta(z_0 \! + \! z_1 \! + \! z_2 \! - \! 1)
	\nonumber
	\\
	& \quad \times
	\int_{\xt_0} \int_{\xt_1} \int_{\xt_2} \int_{\cxt_0} \int_{\cxt_1} \int_{\cxt_2}
	{\cal I}_{\mx}^{(3)} {\cal I}_{\Deltat}^{(3)}
    \,
	4 z_0 z_1 Q^2 \besk_0 \left( Q X_{012}\right) \besk_0 \left( Q X_{\conj{0} \conj{1} \conj{2}}\right)
    \left[ 1 - S_{\contrip}^{\dagger} \right] \left[ 1 - S_{012} \right]
	\nonumber
	\\
	&\qquad
	\times \Bigg\{
	z_1^2
	\Bigg[
	\left(2 z_0 (z_0 + z_2) + z_2^2\right)
	\left(
	\frac{\xt_{20}}{\xt_{20}^2} \cdot
	\left( \frac{\xt_{\conj{2} \conj{0}}}{\xt_{\conj{2} \conj{0}}^2}
	- \half \frac{\xt_{\conj{2} \conj{1}}}{\xt_{\conj{2} \conj{1}}^2} \right)
	- \half \frac{\xt_{\conj{2} \conj{0}} \cdot \xt_{21}}{\xt_{\conj{2} \conj{0}}^2 \xt_{21}^2} \right)
	\nonumber
	\\
	&\qquad \qquad 
	+ \frac{z_2^2}{2}
	\left(
	\frac{\xt_{\conj{2} \conj{0}} \cdot \xt_{21}}{\xt_{\conj{2} \conj{0}}^2 \xt_{21}^2}
	+
	\frac{\xt_{{2} {0}} \cdot \xt_{\conj{2} \conj{1}}}{\xt_{{2} {0}}^2 \xt_{\conj{2} \conj{1}}^2}
	\right)
	\Bigg]
	\nonumber
	\\
	&\qquad \hphantom{\Bigg\{}
	+	
	z_0^2
	\Bigg[
	\left(2 z_1 (z_1 + z_2) + z_2^2\right)
	\left(
	\frac{\xt_{21}}{\xt_{21}^2} \cdot
	\left( \frac{\xt_{\conj{2} \conj{1}}}{\xt_{\conj{2} \conj{1}}^2}
	- \half \frac{\xt_{\conj{2} \conj{0}}}{\xt_{\conj{2} \conj{0}}^2} \right)
	- \half \frac{\xt_{{2} {0}} \cdot \xt_{\conj{2} \conj{1}}}{\xt_{{2} {0}}^2 \xt_{\conj{2} \conj{1}}^2} \right)
	\nonumber
	\\
	&\qquad \qquad 
	+ \frac{z_2^2}{2}
	\left(
	\frac{\xt_{\conj{2} \conj{0}} \cdot \xt_{21}}{\xt_{\conj{2} \conj{0}}^2 \xt_{21}^2}
	+
	\frac{\xt_{{2} {0}} \cdot \xt_{\conj{2} \conj{1}}}{\xt_{{2} {0}}^2 \xt_{\conj{2} \conj{1}}^2}
	\right)
	\Bigg]
	\Bigg\}
	  ,
	\label{eq:ddis-FL-qqbarg-nlo}
\end{align}
for the longitudinal structure function, and
\begin{align}
	\notag
	&\xpom F_{T, \, q \Bar q  g}^{\textrm{D}(4) \, \textrm{NLO}} (\xbj, Q^2, \beta, t)
	=
	2
	\frac{\as \nc \cf Q^4}{\beta} 
	\sum_f e_f^2
	\int_0^1 \frac{\ud z_0}{z_0}
	\int_0^1 \frac{\ud z_1}{z_1}
	\int_0^1 \frac{\ud z_2}{z_2}
	\delta(z_0 \! + \! z_1 \! + \! z_2 \! - \! 1)
	\nonumber
	\\
	& \qquad \times
	\int_{\xt_0} \int_{\xt_1} \int_{\xt_2} \int_{\cxt_0} \int_{\cxt_1} \int_{\cxt_2}
	{\cal I}_{\mx}^{(3)} {\cal I}_{\Deltat}^{(3)}
	\frac{z_0 z_1 Q^2}{X_{012} X_{\conj{0}\conj{1}\conj{2} }}
	\besk_1\left(QX_{012}\right) \besk_1\left(Q X_{\conj{0}\conj{1}\conj{2} }\right)
	\nonumber
	\\
	& \qquad \times
	\Big\lbrace
	\Upsilon^{(|b|^2)}_{\textrm{reg.}} + \Upsilon^{(|c|^2)}_{\textrm{reg.}} 
	+ \Upsilon^{d}_{\textrm{inst.}} + \Upsilon^{e}_{\textrm{inst.}} + \Upsilon^{b \! \times \! c}_{\textrm{interf.}}
	\Big\rbrace
	\left[ 1 - S_{\contrip}^{\dagger} \right] \left[ 1 - S_{012} \right]
	\label{eq:ddis-FT-qqbarg-nlo}
\end{align}
for the transverse structure function. The $\Upsilon$-terms are:
\begingroup
\allowdisplaybreaks
\small
\begin{align}
	\label{eq:ddis-qqbarg-nlo-upsilon-a^2}
	\Upsilon^{(|b|^2)}_{\textrm{reg.}}
	=&
	z_1^2 \Bigg[
		(2z_0(z_0 + z_2) + z_2^2) (1 - 2z_1 (1-z_1))
		\left(\xt_{\conj{0} + \conj{2}; \conj{1}} \cdot \xt_{0+2;1} \right)
		\frac{(\xt_{\conj{2} \conj{0}} \cdot \xt_{{2} {0}})}{\xt_{\conj{2} \conj{0}}^2 \xt_{{2} {0}}^2}
		\nonumber\\
		&
		- z_2 (2z_0 + z_2)(2z_1 - 1)
		\frac{
		    \left( \xt_{\conj{0} + \conj{2}; \conj{1}} \cdot \xt_{\conj{2} \conj{0}} \right)
		    \left( \xt_{{0} + {2}; {1}} \cdot \xt_{{2} {0}} \right)
		    -
		    \left( \xt_{\conj{0} + \conj{2}; \conj{1}} \cdot \xt_{{2} {0}} \right)
		    \left( \xt_{{0} + {2}; {1}} \cdot \xt_{\conj{2} \conj{0}} \right)
		}{\xt_{\conj{2} \conj{0}}^2 \xt_{{2} {0}}^2}
	\Bigg] ,
	\\
	\label{eq:ddis-qqbarg-nlo-upsilon-b^2}
	\Upsilon^{(|c|^2)}_{\textrm{reg.}}
	=&
	z_0^2 \Bigg[
		(2z_1(z_1 + z_2) + z_2^2) (1 - 2z_0 (1-z_0))
		\left(\xt_{\conj{0}; \conj{1} + \conj{2}} \cdot \xt_{0;1+2} \right)
		\frac{(\xt_{\conj{2} \conj{1}} \cdot \xt_{{2} {1}})}{\xt_{\conj{2} \conj{1}}^2 \xt_{{2} {1}}^2}
		\nonumber\\
		&
		- z_2 (2z_1 + z_2)(2z_0 - 1)
		\frac{
		    \left( \xt_{\conj{0}; \conj{1}  + \conj{2}} \cdot \xt_{\conj{2} \conj{1}} \right)
		    \left( \xt_{{0}; {1}  + {2}} \cdot \xt_{{2} {1}} \right)
		    -
		    \left( \xt_{\conj{0}; \conj{1}  + \conj{2}} \cdot \xt_{{2} {1}} \right)
		    \left( \xt_{{0}; {1}  + {2}} \cdot \xt_{\conj{2} \conj{1}} \right)
		}{\xt_{\conj{2} \conj{1}}^2 \xt_{{2} {1}}^2}
	\Bigg] ,
	\\
	\label{eq:ddis-qqbarg-nlo-upsilon-a'}
	\Upsilon^{d}_{\textrm{inst.}}
	=&
	\frac{z_0^2 z_1^2 z_2^2}{(z_0 + z_2)^2}
	- \frac{z_0^2 z_1^3 z_2}{z_0 + z_2}
	\left( 
		\frac{\xt_{{0} + {2}; {1}} \cdot \xt_{{2} {0}}}{\xt_{{2} {0}}^2}
		+
		\frac{\xt_{\conj{0} + \conj{2}; \conj{1}} \cdot \xt_{\conj{2} \conj{0}}}{\xt_{\conj{2} \conj{0}}^2}
		\right)
	\nonumber\\
	&
	+ \frac{z_0^2 z_1 (z_1 + z_2)^2 z_2}{z_0 + z_2}
	\left( 
		\frac{\xt_{{0}; {1}+{2}} \cdot \xt_{{2} {1}}}{\xt_{{2} {1}}^2}
		+
		\frac{\xt_{\conj{0}; \conj{1}+ \conj{2}} \cdot \xt_{\conj{2} \conj{1}}}{\xt_{\conj{2} \conj{1}}^2}
		\right) ,
	\\
	\label{eq:ddis-qqbarg-nlo-upsilon-b'}
	\Upsilon^{e}_{\textrm{inst.}}
	=&
	\frac{z_0^2 z_1^2 z_2^2}{(z_1 + z_2)^2}
	- \frac{z_0 z_1^2 (z_0 + z_2)^2 z_2}{z_1 + z_2}
	\left( 
		\frac{\xt_{{0} + {2}; {1}} \cdot \xt_{{2} {0}}}{\xt_{{2} {0}}^2}
		+
		\frac{\xt_{\conj{0} + \conj{2}; \conj{1}} \cdot \xt_{\conj{2} \conj{0}}}{\xt_{\conj{2} \conj{0}}^2}
		\right)
	\nonumber\\
	&
	+ \frac{z_0^3 z_1^2 z_2}{z_1 + z_2}
	\left( 
		\frac{\xt_{{0}; {1}+{2}} \cdot \xt_{{2} {1}}}{\xt_{{2} {1}}^2}
		+
		\frac{\xt_{\conj{0}; \conj{1}+ \conj{2}} \cdot \xt_{\conj{2} \conj{1}}}{\xt_{\conj{2} \conj{1}}^2}
		\right) ,
	\\
	\label{eq:ddis-qqbarg-nlo-upsilon-ab}
	\Upsilon^{b \! \times \! c}_{\textrm{interf.}}
	=&
	- z_0 z_1
	\left[ z_1(z_0 + z_2) + z_0(z_1 + z_2) \right]
	\left[ z_0(z_0 + z_2) + z_1(z_1 + z_2) \right]
	\nonumber\\
	&\quad \times \!
	\left[
		\left( \xt_{\conj{0} + \conj{2}; \conj{1}} \cdot \xt_{0;1+2} \right)
		\frac{(\xt_{\conj{2} \conj{0}} \cdot \xt_{{2} {1}})}{\xt_{\conj{2} \conj{0}}^2 \xt_{{2} {1}}^2}
		+
		\left( \xt_{\conj{0}; \conj{1} + \conj{2}} \cdot \xt_{0+2;1} \right)
		\frac{(\xt_{\conj{2} \conj{1}} \cdot \xt_{{2} {0}})}{\xt_{\conj{2} \conj{1}}^2 \xt_{{2} {0}}^2}
	\right]
	\nonumber\\
	&
	+ z_0 z_1 z_2 (z_0-z_1)^2
	\nonumber\\
	& \quad \times \!
	\Bigg[
		\frac{
		    \left( \xt_{\conj{0} + \conj{2}; \conj{1}} \cdot \xt_{\conj{2} \conj{0}} \right)
		    \left( \xt_{{0}; {1}  + {2}} \cdot \xt_{{2} {1}} \right)
		    -
		    \left( \xt_{\conj{0} + \conj{2}; \conj{1}} \cdot \xt_{{2} {1}} \right)
		    \left( \xt_{{0}; {1}  + {2}} \cdot \xt_{\conj{2} \conj{0}} \right)
		}{\xt_{\conj{2} \conj{0}}^2 \xt_{{2} {1}}^2}
    	\nonumber\\
    	& \quad \quad
		+
		\frac{
		    \left( \xt_{\conj{0}; \conj{1}  + \conj{2}} \cdot \xt_{\conj{2} \conj{1}} \right)
		    \left( \xt_{{0} + {2}; {1}} \cdot \xt_{{2} {0}} \right)
		    -
		    \left( \xt_{\conj{0}; \conj{1}  + \conj{2}} \cdot \xt_{{2} {0}} \right)
		    \left( \xt_{{0} + {2}; {1}} \cdot \xt_{\conj{2} \conj{1}} \right)
		}{\xt_{\conj{2} \conj{1}}^2 \xt_{{2} {0}}^2}
	\Bigg] \!.
\end{align}%
\endgroup
Both diffractive structure functions factorize quite neatly: the dipole scattering amplitudes, the final state transfer functions, and the virtual photon LFWF are factorized into their own parts. A more detailed discussion of the results can be found in Ref.~\cite{Beuf:2022kyp}.

\section{Recovering previous results as limiting cases}

A natural cross check for our new results is to compare in detail with previously known results for $F_T^D$. A complication for performing such a comparison is the fact that the previous results were derived in specific kinematical limits that make them challenging to reverse engineer from our final results, especially so for the large-$Q^2$ result. Detailed calculations can be found in Ref. \cite{Beuf:2022kyp}.

Of the two known results to compare with, the derivation of the large-$\mx$ result turns out to be more straightforward, however, it requires the inclusion of the contributions of the diagrams where the gluon is emitted into the diffractive system in the final state. This is necessary since in the large-$\mx$ limit the tree-level diagrams alone become divergent, which is canceled by the inclusion of the final state emissions. For this calculation, we derived the final state emission contribution only in the large-$\mx$ limit leveraging the property of the virtual photon LCWF that in the $z_2 \to 0$ limit the $\gamma^* \to q \bar q$ and $q \to qg $ contributions can be factorized.

In order to be able to recover the large-$Q^2$ result for the $\qqbg$ contribution to $F_T^D$, it was necessary to understand that the original result was derived in a picture parametrized in minus-momentum fractions of the target, instead of the plus-momentum picture that we use. It was necessary to write the full NLO result for $F_T^D$  in full momentum space, after which it was possible to translate between the two momentum pictures, and apply the kinematical approximations used in the original work. We were able to verify the original result explicitly from first principles, and the result has interesting features considering the derivation starting from the dipole picture. The large $\log Q^2$ is made explicit in the result, and the DGLAP $g \to q \bar q$ splitting function emerges from the squared virtual photon wavefunction in the aligned jet limit.

\section{Conclusions}

We have calculated the next-to-leading order $q \bar q g$ contribution to the diffractive DIS cross sections for both the longitudinal and transverse virtual photon using the CGC formalism. This provides the first NLO accuracy result for the diffractive longitudinal structure function, and a fundamental theory accuracy improvement for the transverse structure function. As a cross check, we recovered two well-known results for the $\qqbg$ contribution as limiting cases of our full result. The numerical implementation of these results is underway, and once finished can be applied to EIC phenomenology, and also to learn more about saturation due to the enhanced sensitivity of the diffractive structure functions.

\vspace{2mm}
{
\footnotesize
\textit{Acknowledgements}
T.L and H.M are supported by the Academy of Finland, the Centre of Excellence in Quark Matter (project 346324) and projects 338263, 346567 and 321840.
H.H is supported by the Academy of Finland, the Centre of Excellence of Inverse Modelling and Imaging.
G.B is supported in part by the National Science Centre (Poland) under the research grant no. 2020/38/E/ST2/00122 (SONATA BIS 10).
Y.M acknowledges financial support from Xunta de Galicia (Centro singular de investigaci\'on de Galicia accreditation 2019-2022); the ``Mar\'{\i}a de Maeztu'' Units of Excellence program MDM2016-0692 and the Spanish Research State Agency under project PID2020-119632GB-I00; European Union ERDF.
G.B and Y.M acknowledge financial support from MSCA RISE 823947 ”Heavy ion collisions: collectivity and precision in saturation physics” (HIEIC).
This work was also supported under the European Union’s Horizon 2020 research and innovation programme by the European Research Council (ERC, grant agreement No. ERC-2018-ADG-835105 YoctoLHC) and by the STRONG-2020 project (grant agreement No. 824093).
The content of this article does not reflect the official opinion of the European Union and responsibility for the information and views expressed therein lies entirely with the authors. 
}
    
\bibliographystyle{JHEP-2modM.bst}
\bibliography{refs}

\end{document}